\documentclass[aps,prl,reprint,groupedaddress,twocolumn]{revtex4-1}
\usepackage{graphicx}
\usepackage{dcolumn}
\usepackage{color}
\usepackage{tikz}
\usetikzlibrary{shapes}
\usepackage{gensymb}
\usepackage{epstopdf}
\usepackage{mathtools}
\usepackage{microtype}
\begin{document}
\preprint{AIP/123-QED}
\title[Simple Model]{Quantum Tunneling Theory of Cooper Pairs as Bosonic Particles}
\vspace{1cm}
\author{Edgar J. Pati\~no}
\affiliation{Departamento de F{\'i}sica, Grupo de F\'isica de la Materia Condensada, Universidad de los Andes, Carrera 1 No. 18A-12, A.A. 4976-12340, Bogot{\'a}, Colombia.}
\affiliation{School of Physical Sciences and Nanotechnology, Yachay Tech University, Urcuqu\'i, Ecuador}
\
\author{Daniel  Lozano-G\'omez}
\affiliation{Departamento de F{\'i}sica, Grupo de F\'isica de la Materia Condensada, Universidad de los Andes, Carrera 1 No. 18A-12, A.A. 4976-12340, Bogot{\'a}, Colombia.}
\affiliation{Department  of  Physics  and  Astronomy,  University  of  Waterloo,  Ontario,  N2L  3G1,  Canada}
%
%
\date{\today}
\begin{abstract}

We propose a simple phenomenological theory for quantum tunneling of Cooper pairs based on their boson like nature.
Thus it applies in the absence of quasiparticle excitations (fermions), and should be suitable for boson like particles at the low energy regime. Around zero bias voltage our model reveals a rapid increase in tunneling current which rapidly saturates. This manifests as a zero bias conductance peak that strongly depends on the superconductors transition temperature. This low energy tunneling of Cooper pairs could serve as an alternative explanation for a number of tunneling experiments where zero bias conductance peak has been observed.
%
%
\end{abstract}
\pacs{74.25.Ha, 74.45.+c, 75.30.Et}
%
\maketitle

\textbf{I. INTRODUCTION}\\\\
%
%
Quantum tunneling in solid state devices has been a subject of intensive research since middle of the 20th century after its original proposal in quantum mechanics theory. It has been widely investigated both in theory and experiments in a number of branches of physics including; atomic physics for explaining the decay of a nucleus \cite{Sexl1929}, cosmological physics for study of thermal emission black holes \cite{Hawking1975,Kraus1995}, Rb atoms Bose Einstein condensates in bosonic Josephson junctions \cite{Albiez2005} and 
solid state physics \cite{Kontos2001,Patino2015, Kelkar2017} where artificially fabricated tunnel devices have been experimentally realized with multiple applications. Indeed, when two electrodes are separated by thin insulating material, single electrons can tunnel across the barrier as demonstrated by numerous experiments even at room temperature \cite{Patino2015, Kelkar2017}.  One of the most popular models for tunnel junctions for electrons was proposed by Simmons in 1963 \cite {Simmons1}, and describes tunneling of fermionic current through a junction which is temperature dependent. This model has been widely used to explain results obtained from quantum tunneling devices and is the inspiration for the work we present here. 

On the other hand, for paired  electrons forming Cooper pairs in superconductor/insultator/superconductor (S/I/S) junctions, tunneling effects were predicted by B. D. Josephson in 1962 and confirmed experimentally soon after. 
Indeed, the  DC Josephson effect predicted a dc current at zero bias voltage, and has been confirmed by enormous amount of experiments. 
However, at T=0 if the voltage is increased, no current can flow from one electrode to the other until it reaches $eV = 2\Delta$ i.e. the energy needed to break Cooper pairs that leads to quasiparticle current. This is not true for N/I/S junctions where for $T>0$ the energy of quasiparticle excitations, from the normal metal,  allows tunneling at lower voltages via the so called Andrev reflection processes,  not relevant for the present analysis. For (S/I/S) junctions at finite temperatures when phonon energies 
and voltages $> 2\Delta$, a fraction of Cooper pairs break into quasiparticles producing so called subgap or excess currents.  Given that in this case quasiparticles are made of electrons, tunneling processes are of fermionic nature
leading to tunneling current-voltage characteristics of similar shape to the ones obtained 
for normal tunneling junctions \cite{Patino2015} i.e a indefinitely increasing function of voltage without saturation value. 
The main problem is that sometimes these currents are observed experimentally at temperatures and voltages below  $2\Delta$  \cite{Im2010,Schmidt2012} lower than the energy required for thermally induced excess currents without a clear explanation on its origin. In other experiments on Nb based S-I-S junctions \cite{1480930,Dettmann1979}, where considerably higher currents than anticipated from BCS theory where found, the explanation was attributed to the existence of  a normal conducting layer on the superconducting film \cite{Seidel1980,Blonder1982}. Zero bias conductance peak was also been observed in S-I-S junctions \cite{Doring2014}
by tuning the barrier thickness. These results where  also explained by normal layer formation leading to  Andreev- reflection processes. In all these experiments there was not direct experimental evidence of this normal layer. 
More recently experimental evidence  of non uniform boson distribution 
in Nb-AlOx-Nb Josephson junctions arrays was reported \cite{Silvestrini2007,Ottaviani2014, Lorenzo2014}. This was attributed to Bose-Einstein condensation and boson hopping  \cite{Ottaviani2014, Lorenzo2014}.  
between superconducting islands of the array. 
In this letter, based on the assumption of Cooper pairs behaving as bosons, we derive a theory for the current flow  
of bosons across a S/I/S tunnel junction. 
This proposes a different way to transfer Cooper pairs as
boson like particles, in addition to quasiparticle tunneling. This could take place experimentally for voltages and temperatures bellow the energy gap where small excess currents are observed although quasi particle excitations are absent. Once the voltage and temperature 
reach $\sim$2$\Delta$, boson currents should be replaced by quasi particle currents.
\\
\\
\textbf{II. TUNNELING OF BOSONS}\\
\\
Lets start by considering two boson reservoir separated by a insulating barrier material. In each reservoir, bosons are assumed to have intrinsic charge and share the same energy state at the chemical potential $\mu$ with a probability density distribution given by the Bose-Einstein (BE) distribution. The probability  $D(E_x)$ that an incident boson, with kinetic energy $E_{x}=m v_{x}^{2}/2$  component along the $x$ direction, crosses the potential barrier $V(x)$  
can be described by means of the WKB approximation
\begin{eqnarray}\label{eq:D}
D(E_x)=\exp\left(-2\int_{x_1}^{x_2}\sqrt{ \frac{2m\ast}{\hbar^2}\left[V(x)-E_x\right]} dx\right),
\end{eqnarray}
where, bosons are subject to a potential energy $V(x)=\mu+\phi(x)$ dictated by the barrier height $\phi(x)$, defined in the interval $[x_1,x_2]$, and chemical potential $\mu$, close to zero at low temperatures.    
This allows to calculate the number of bosons tunneling from the left side (N$_1$) and right side (N$_2$);
\begin{eqnarray}\label{eq:n1}
N_1&=&\frac{4\pi m^{\ast }}{h^3}\int_0^{E_m} D(E_x)\left[\int_0^\infty B(E)dE_r\right]dE_x \\
\nonumber
N_2&=&\frac{4\pi m^{\ast}}{h^3}\int_0^{E_m} D(E_x)\left[\int_0^\infty B(E+eV)dE_r\right]dE_x,
\end{eqnarray}
where $E_m$ is the maximum energy of the electrons and the integral has been expressed in polar coordinates i.e. $E_r = E_y + E_z$.  
Through the expression above, a net bosonic current flow (J$_B$) is obtained from the difference between these two values;
\begin{eqnarray}\label{eq:JB}
J_B&=e^\ast&(N_1-N_2) 
\end{eqnarray}
which can be rewritten as;
\begin{widetext}\label{eq:JB2}
\begin{eqnarray}\label{eq:JB2}
J_B&=&\frac{4\pi m^{\ast}e^\ast}{h^3}\int_0^{E_m}D(E_x)\left[\int_0^\infty B(E)dE_r-B(E+eV)dE_r\right] dE_x.
\end{eqnarray}
\end{widetext}
In the present work, we consider a fixed rectangular potential barrier $\phi_0$ where the barrier height is approximately the same for the studied voltage regime. This assumption is consistent with the low voltage approximation ($V\approx$0 as in  \cite {Simmons1})
where barrier shape does not change upon applied voltage. This is justified when considering superconductors as bosons reservoirs as we shall see in the next section.
Therefore the potential takes a constant value of  $V (x)=\mu+\phi_0$,  
%
$\mu$  the highest occupied energy level and $\phi_0$ the barrier height; the tunneling probability simplifies to;
\begin{widetext}
\label{eq:D_rec}
\begin{eqnarray}\label{eq:D_rec}
D(E_x)&=&\exp\left(-2\int_{x_1}^{x_2}\sqrt{ \frac{2m\ast}{\hbar^2}\left[\phi_0+\mu-E_x\right]} dx\right),
\end{eqnarray}
\end{widetext}
Integrating equation \ref{eq:D_rec} and the approximation used by \cite {Simmons1} gives the solution 
\begin{widetext}\label{eq:D_rec_S}
\begin{eqnarray}\label{eq:D_rec_S}
D(E_x)&\approx &\exp\left(\frac{-2(2m^\ast)^{1/2}}{\hbar}\beta s\left[\mu+\phi_0-E_x\right]^{1/2} \right),
\end{eqnarray}
\end{widetext}
where $s=x_2-x_1$ corresponds to the barrier width and $\beta$ is a correction factor which
can be chosen to be unity for the low voltage regime \cite{Simmons1}.
%
Finally, the temperature dependence is included in the difference of BE distributions integral (BEI); 
%
\begin{widetext}\label{eq:JB2Temp}
\begin{eqnarray}\label{eq:JB2Temp}
\mathrm{BEI}=\left[\int_0^\infty B(E)dE_r-B(E+eV)dE_r\right] = 
\int_0^{\infty} \left[\frac{1}{e^{(E-\mu)/k_B T}-1}-\frac{1}{e^{(E+eV-\mu)/k_B T}-1}\right]dE_r
\nonumber 
\\
= \ln\left[\frac{1-e^{(\mu-E_x-eV)/k_B T} }{1-e^{(\mu-E_x)/k_B T}}\right],
\end{eqnarray}
\end{widetext}

when solved  \cite{derivative} 
%

gives the final relation between current density and voltage across the leads;
\begin{widetext}\label{eq:JB}
\begin{eqnarray}\label{eq:JB}
J_B
&=&\frac{4\pi m^{\ast }e^\ast}{h^3} \int_0^{E_m} \exp\left(\frac{-2(2m^\ast)^{1/2}}{\hbar}\beta s\left[\mu+\phi_0-E_x\right]^{1/2} \right)  \ln\left[\frac{1-e^{(\mu-E_x-eV)/k_B T} }{1-e^{(\mu-E_x)/k_B T}}\right]dE_x.\nonumber\\
\end{eqnarray}
\end{widetext}
\textbf{III. TUNNELING OF COOPER PAIRS AS BOSONIC PARTICLES}\\

\subsection{The superconductor case}
\begin{figure}[!ht]
\centering
\includegraphics[width=0.5\textwidth]{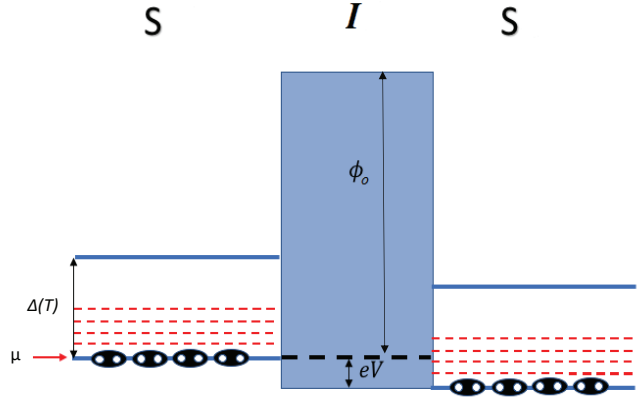}
%
\caption{Insulating potential barrier between two superconductors (boson reservoirs). Here Cooper pairs remain as bound particles in the ground state at energies bellow the energy gap $\Delta(T)$ and barrier height $\phi_{o}$.}
\label{barrier}
\end{figure}

Consider a superconducting junction 
where Cooper pairs are assumed to behave as bosons \cite{samuelsson,fujita, dellano,Mamedov2007}
and thus obey Bose-Einstein statistics. This is possible for a composite of even fermions with finite center of mass momentum \cite{dellano,fujita}. 
Thus Cooper pairs occupy a single energy level $\mu$, at $T=0$, that corresponds to the ground state energy of the system. This energy level has the highest density of states and is separated from the quasiparticle energy states by the energy gap $ \Delta_0 $. 
On the other hand, for non zero temperature, the BE distribution will allow higher energy states to be occupied by the bosons (Cooper pairs) in the vicinity of the ground state.
In this scenario, illustrated in Fig. \ref{barrier}, each electrode can be considered a Cooper pair/boson reservoir separated by a barrier material. 
Due to their superconducting properties, the number of bosons in each reservoir depend on temperature.
This leads to reduction of Cooper pair density with increasing temperature up to $T_{c}$ where it drops to zero.
 We will start from eq. \ref{eq:D_rec_S} where we considered the case of a rectangular potential barrier of height $\phi_{0}$ and width $s$ valid for the low voltage regime.  Indeed, having the requirement of voltages $V<$  $\Delta_0 $, that prevents Cooper pairs from breaking, given that $\Delta_0 $ is typically of the order of millivolts and the barrier height $\phi_{0}$ of the order of eV, justifies the  low voltage approximation used here where  $V< \Delta_0<< \phi_{0}$.
For simplicity we consider $\mu=0 $, as before we take the correction factor ($\beta=1$) and  $e^\ast = 2e$ and $ m^\ast = 2m$ due to the intrinsic nature of bosons in our condensate  made up of paired of electrons.  Thus the tunneling probability (eq. \ref {eq:D_rec}) and current density (eq. \ref {eq:D_rec}) can be expressed as 
\begin{widetext}
\begin{eqnarray}
D(E_x)=\exp\left(\frac{-2(4m\phi_0)^{1/2}}{\hbar} s \left[1-\frac{E_x}{\phi_0}\right]^{1/2} \right)
\end{eqnarray}
\end{widetext}
\begin{widetext}\label{eq:JB_E}
\begin{eqnarray}\label{eq:JB_E}
J_B=\frac{16\pi me k_BT }{h^3} \int_0^{E_m} \exp\left(\frac{-2(4m\phi_0)^{1/2}}{\hbar} s \left[1-\frac{E_x}{\phi_0}\right]^{1/2} \right)  \ln\left[\frac{1-e^{(-E_x-eV)/k_B T} }{1-e^{-E_x/k_B T}}\right]dE_x.
\end{eqnarray}
\end{widetext}
%
The last integral is an almost exact expression. Finally, writing the constant terms as 
$A = 2(4m)^{1/2}\hbar^{-1}  s$,  $C_1=\left(\frac{16\pi m e k_B^2T^2}{h^3 }\right) \exp\left(-A\phi_0^{1/2}\right)$, using the binomial approximation in the transmission coefficient  in equation \ref {eq:JB_E} and the substitution;  $u=E_x/k_BT$ the expression for the bosonic current is simplified to:
\begin{widetext}
\begin{eqnarray}
J_B&=&C_1 \int_0^{\frac{Em}{k_BT}} \exp\left( k_B T\frac{A}{2\phi_0^{1/2} } u
\right)  \ln\left[ \frac{1-e^{[-eV]/k_BT } }{1- e^{-u}}\right]du.\nonumber
\end{eqnarray}
\end{widetext}
This equation is similar to the  one found by Simmons \cite {Simmons2}, with a different constant value $C_1$, potential used and  signs in the logarithm.
We know that the maximum  energy $E_m$ is much greater than the thermal energy that the system can reach thus $ E_m /k_BT\gg 1$, this condition let us 
make an extension to the integral such that;
\begin{widetext}\label{eq:integral}
\begin{eqnarray}\label{eq:integral}
J=\frac{C_1}{k_BT} \int_0^{Em} \exp\left( \frac{A}{2\phi_0^{1/2} }E_x \right)  \ln\left[\frac{1-e^{-eV-E_x/k_B T }}{1- e^{-E_x/k_BT}}\right]dE_x.\nonumber
\end{eqnarray}
\end{widetext}
Using the constants
\begin{eqnarray}
C_1&=&\left(\frac{16\pi m e k_B^2T^2}{h^3 }\right) \exp\left(-A\phi_0^{1/2}\right)\nonumber\\
C_2&=&  k_BT\frac{A}{2\phi_0^{1/2} }\nonumber\\
C_3&=& e^{-eV/k_BT },\nonumber\\
\end{eqnarray}
a new equation, which should be integrated, is obtained:
\begin{widetext}\label{eq:integral}
\begin{eqnarray}\label{eq:integral}
J&=&C_1 \int_0^{\infty} \exp\left(C_2 (u-\frac{1}{k_BT})\right)  \ln\left[ \frac{1-C_3 e^{-u} }{1- e^{-u}}\right]du.
\end{eqnarray}
\end{widetext}
That may not have an analytical solution, therefore one may benefit of the approximation used in \cite{Murphy} to simplify the logarithm, for currents caused by thermo-ionic effects. 
The approximation in \cite{Murphy} can be applied  when the energy is greater than the chemical potential,  $\mu$, plus a small contribution of  $ k_B T $.
\begin{eqnarray}
J&\simeq&\frac{ C_1  }{1-C_2}  (1-e^{-eV/k_BT}).
\end{eqnarray}

Which, replacing the constant values leaves us with;
\begin{eqnarray}
J&\simeq&\frac{ 32\pi m^2 e k_B^2T^2\phi_0^{1/2} }{2\phi_0^{1/2}h^3 -  h^3 k_BT A }e^{-A\phi_0^{1/2}} (1-e^{-eV/k_BT})
\end{eqnarray}
\subsection{Varying the boson density by adding the occupation number}

\begin{figure}[!ht]
\centering
\includegraphics[width=0.50\textwidth]{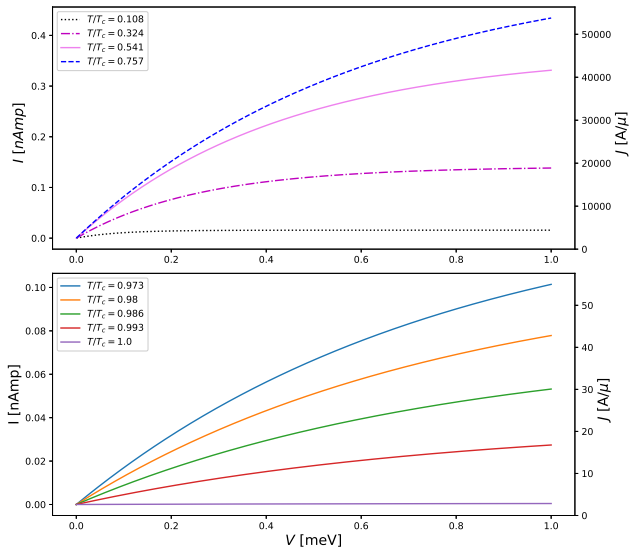}
%
%
%
\caption{I-V characteristics obtained from equation \ref{eq:integral} for Nb/Al$_{2}$O$_{3}$/Nb junctions at low and high temperatures bellow $T_{c}$.} 
\label{fig:IV_D}
\end{figure}

In the model presented so far we have ignored  the occupation number part of the BE distribution, normally assumed to be constant.  
However for the superconductor case the number of bosons  $N$ in the reservoirs is not fixed and strongly depends upon temperature since the Cooper pairs will increase as the temperature drops. 
In the present model the temperature dependency of $N$ is taken into account by multiplying the occupation number, $N$, in the BE distribution or equivalently by the current density;

\begin{eqnarray}
J \to N(T)\times J \nonumber
\end{eqnarray}
The explicit temperature dependence of $N(T)$ can be obtained from the  
the following phenomenological expression for the penetration length $\lambda$  \cite{Schmidt2}; 
\begin{widetext}
\begin{eqnarray}
\lambda&=&\frac{\lambda(0)}{\left(1-[T/T_c]^4\right)^{\frac{1}{2}}}=\left(\frac{mc^2}{4\pi N e^2}\right)^{1/2}, \nonumber
\end{eqnarray}
\end{widetext}

where $\lambda(0)$ is the penetration length at absolute zero. 
Isolating $N(T)$ from the penetration depth expression gives its temperature dependence;

\begin{eqnarray}\label{occupancy}
N(T)=\frac{mc^2}{4\pi \lambda^2(0)e^2}\left[  1-\left(\frac{T}{T_c}\right)^4 \right],
\end{eqnarray}
which as expected goes to $0$ at the critical temperature. This allows us to obtain the expression for current density of superconducting tunnel junctions; 
%
\begin{widetext}
\begin{eqnarray}\label{eq:integral}
%
J&\simeq& \frac{8m^3c^2  k_B^2\phi_0^{1/2}T^2}{\lambda^2(0)e  h^3  ( 2\phi_0^{1/2} -  k_B A T)}\left[  1- \left(\frac{T}{ T_c}\right)^4 \right]   e^{-A\phi_0^{1/2}} (1-e^{-eV/k_BT}).
\end{eqnarray}
\end{widetext}
%
%
In order to have better insight of equation \ref{eq:integral} we use the junction parameters 
for Aluminium oxide  $Al_{2}$$O_{3}$ a customary barrier material
obtained from previous experiments \cite{Patino2015}. Here tunnel junctions with barrier height $\phi=1.8$ eV  and width $s =2.079$ nm where produced with a barrier cross section area A = 346 $\mu$m $\times$ 375 $\mu$m. Regarding superconductor parameters, the values for Niobium of  $\lambda(0)$=0.047 nm and  $T_{c}$=9.25 K  have been used.
Replacing these parameters in expression \ref{eq:integral} the I-V and J-V characteristic curves are plotted in Fig. \ref{fig:IV_D} for several temperatures. 

\begin{figure}[!ht]
\centering
\includegraphics[width=0.5\textwidth]{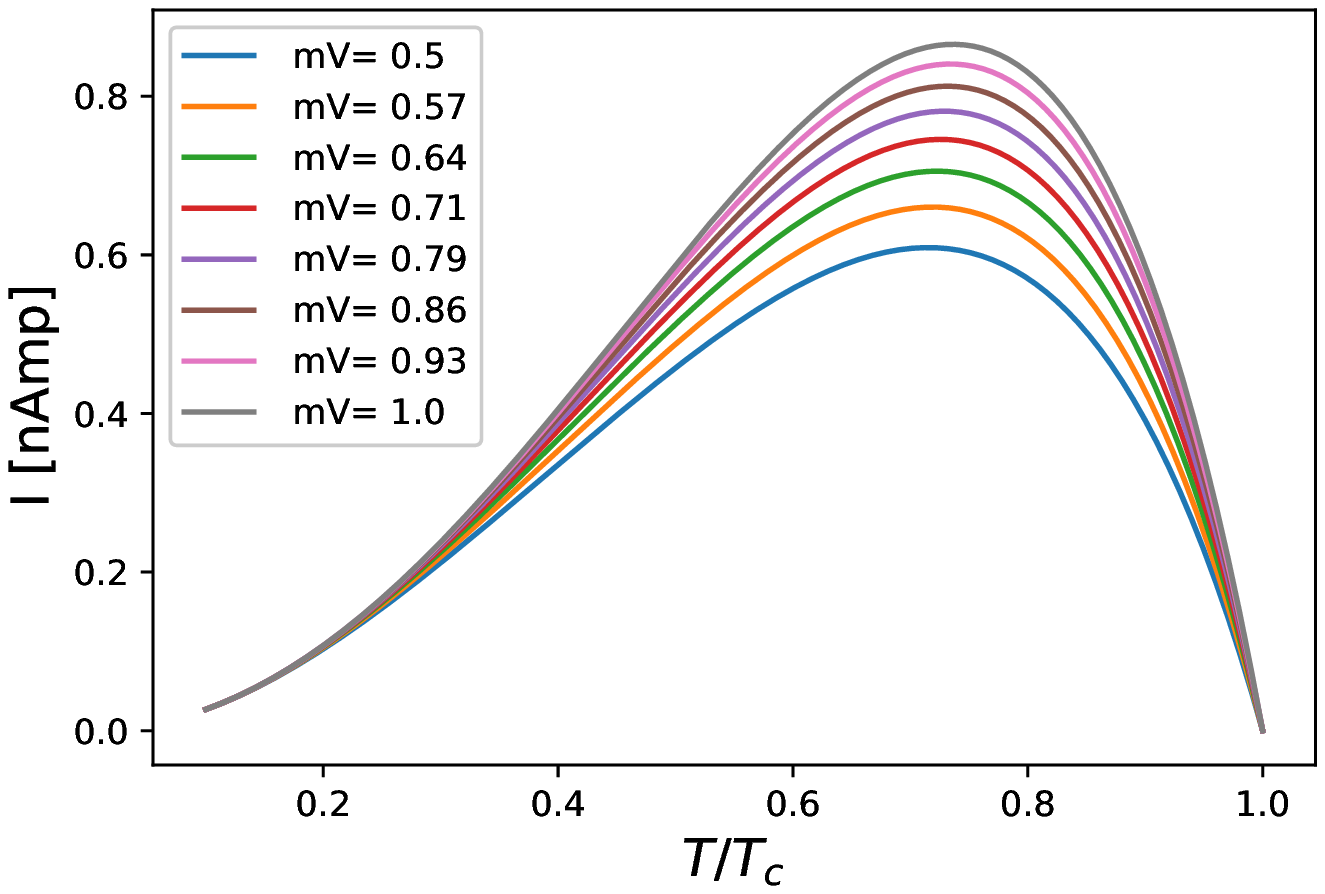}
\centering
\includegraphics[width=0.4\textwidth]{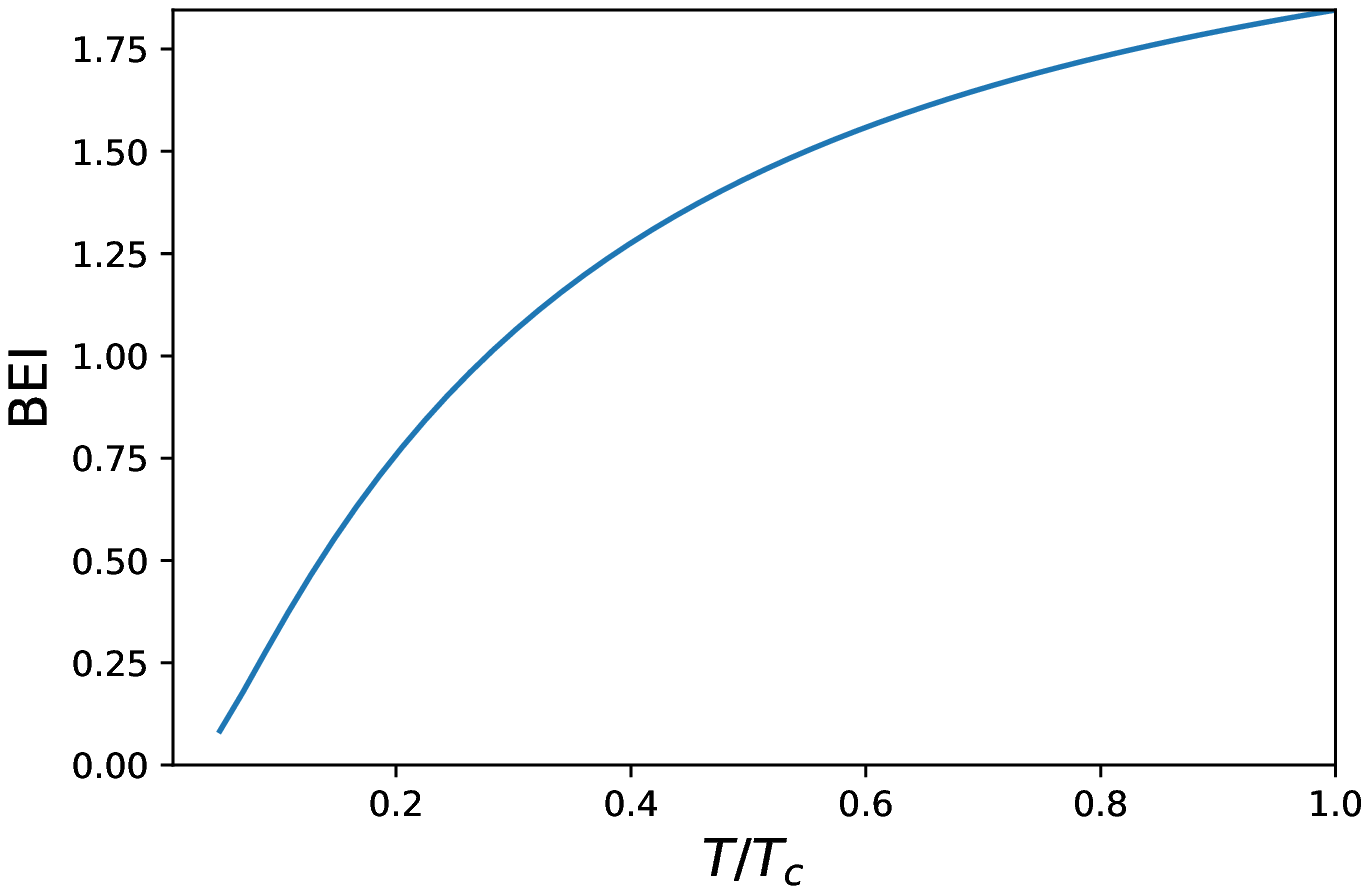}
\caption{Upper figure shows Nb/Al$_{2}$O$_{3}$/Nb junction current vs temperature for set of fixed voltages. Lower figure shows the Bose Einstein distribution difference integral (Eq. \ref{eq:JB2Temp})  vs temperature.}
\label{I_T}
\end{figure}

It is interesting to notice the non monotonic behavior of the current at low and high temperatures. At low temperatures the I-V characteristics show an  increase in the values of current as temperature rises while at high temperatures there is a sharp reduction in the current values as temperature approaches to $T_{c}$. This effect can be more clearly seen by fixing the voltage and plotting the current as function of temperature as depicted in Fig. \ref{I_T}. Starting at zero temperature the rise in tunneling with temperature, for temperatures T/Tc $<$0.7,  can be understood by analyzing the the integral of the  Bose Einstein distribution difference between reservoirs given in eq. \ref{eq:JB2Temp}. Such expression, with a logarithmic  dependance, represents the net difference in number of bosons between reservoirs at a given junction voltage and temperature.  
Using constant reference values of voltage $V$=1 $mV$ and $E_{x}=$ $1\times 10^{-4}$ eV it is possible to obtain such difference solely as function of temperature. This is shown at Inset Fig. \ref{I_T} where a significant increase between zero and transition temperatures is observed. This indicates and important broadening of the BE distribution at finite temperatures that leads to the increase of boson occupancy probability at higher energy levels above the chemical potential and thus enhancing the transmission probability across the barrier for temperatures T/Tc $<$0.7.

\begin{figure}[!ht]
%
\centering
\includegraphics[width=0.5\textwidth]{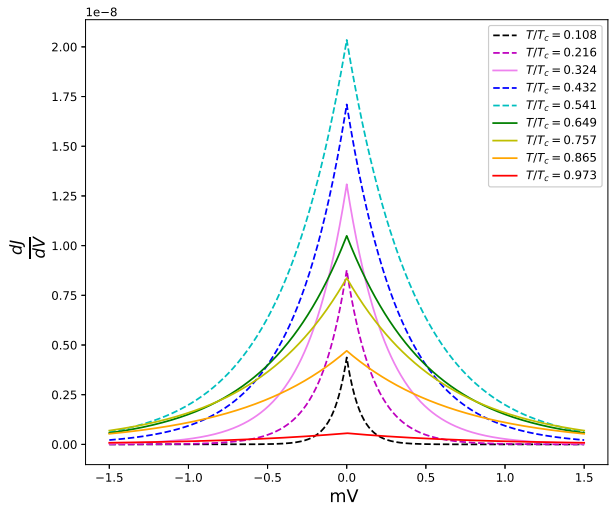}
%
\caption{Zero bias conductance peak as function of temperature obtained from the I-V characteristics}
\label{fig:IT_mod}
\end{figure}

On the other hand  for T/T$_{c}>$0.7  the reduction of tunneling current with  temperature can be explained as being the result of the reduction in the number of Cooper pairs, expressed in eq. \ref{occupancy},  as temperature approaches the phase transition. 
%
Finally taking the derivative of the current we obtain the conductance as function of voltage as plotted in Fig. \ref{fig:IT_mod}. This reveals a zero value conductance peak with the same temperature dependence as the IV characteristics described earlier. 
This peak could serve as an alternative explanation to a number of experimental results found in the literature to be discussed in the next section.
\\ 
 \\
\textbf{IV. SUMMARY AND DISCUSSION}\\ 
\\
	An expression for tunneling of Cooper pairs that follows Bose-Einstein statistics 	
has been obtained in Eq. \ref{eq:integral}. This equation gives the IV characteristics for S/I/S symmetric tunnel junctions as function of temperature. For the assumption of Cooper pairs behaving as bosons, to be feasible the energy from the voltage across the junction ($eV$) plus thermal energy ($k_{b}$$T$) should be kept below twice the energy gap i.e. $eV$ and $k_{b}$ $T$$<2\Delta$. 
Indeed, in such regime quasiparticle excitations can not take place preventing single electron's tunneling. Given the low applied voltage restriction compared to the barrier height where $\phi_{o}$ $>$$>$eV, a constant rectangular barrier can be chosen as good approximation. 
As opposed to quasiparticle tunneling,  bosons tunneling shows extremely low values of current 
and the IV characteristics approach an asymptotic value as the voltage increases. This can be explained due to the finite number of Cooper pairs available at the superconductors which restricts the boson current. 	

Regarding the temperature dependence of tunneling current seen in Fig.  \ref{I_T} it is interesting to note its non monotonic behaviour.  
Around zero temperature, the tunneling current increases with temperature until it reaches a maximum value around 0.7  $T/T_{c}$. This can be explained as result of the rise in tunneling probability, due to broadening of the BE distribution with increasing temperature. The maximum value is then followed by a reduction of tunneling current due to the drop in Cooper pair density, as temperature increases towards $T_{c}$.
Finally by taking the derivative of the I-V characteristics the conductance in Fig.  \ref{fig:IT_mod} shows a zero bias conductance peak with its maximum around  0.7  $T/T_{c}$ and vanishes around zero or  $T_{c}$.  It is interesting to notice the zero bias conductance peak has been obtained without the need of a thin metallic layer used in the explanation of a number or previous experiments \cite{Doring2014}. As follows from this theory this peak should be taken as the finger print of boson tunneling in other S/I/S systems where thin metallic layers can be ruled out.  

%

%

\textbf{V CONCLUSION}\\ 

In previous theories of transport in S/I/S superconducting junctions,  quasiparticle electron tunneling has been considered as responsible for the quantum tunneling currents,  not considering boson like behavior of Cooper pairs. 
Here we proposed a simple theory for quantum tunneling of Cooper pairs that exclusively follows from their boson like nature. It should only apply when the applied voltage and temperature is below twice the energy gap
i.e. in the absence of quasiparticle excitations. Around zero bias voltage, our model predicts a zero bias conductance peak that strongly depends on the superconductor's temperature. 
This boson tunneling theory offers a possible explanation for a number of tunneling experiments where subgap currents appear that may or not include zero bias conductance peak that varies with temperature. Also may shed light into experiments of Josephson junctions arrays when Bose-Einstein condensation is believed explain their results.
\section{ACKNOWLEDGEMENT}
We would like to acknowledge D. Chevallier  for  useful  discussions.  This  work  was  partially funded  by Convocatoria Programas 2012 Vicerrector\'ia de Investigaciones of Universidad de los Andes Bogot\'a, Colombia.
%
%
\section*{References}
\bibliography{references}
\end{document}